\documentclass{ws-fnl}
\usepackage{cite}

\usepackage{amsmath,bbm}

\catcode`\|=\active \def|{
\fontencoding{T1}\selectfont\symbol{124}\fontencoding{\encodingdefault}}
\newcommand{\mathd}{\mathrm{d}}
\newcommand{\mathe}{\mathrm{e}}
\newcommand{\nocomma}{}
\newcommand{\nosymbol}{}
\newcommand{\tmem}[1]{{\em #1\/}}
\newcommand{\tmmathbf}[1]{\ensuremath{\boldsymbol{#1}}}
\newcommand{\tmop}[1]{\ensuremath{\operatorname{#1}}}
\newcommand{\tmtextbf}[1]{{\bfseries{#1}}}

\begin{document}

\markboth{Vacchini}{QCFN Special Issue: Decoherence and noise in open quantum system dynamics}

\catchline{}{}{}{}{}

\title{Decoherence and noise in open quantum system dynamics}

\author{\footnotesize BASSANO VACCHINI}

\address{Dipartimento di Fisica, Universit\`a degli Studi di Milano,
Via Celoria 16\\ Milan, 20133, Italy\\
INFN, Sezione di Milano, Via Celoria\\ Milan, 20133, Italy\\
bassano.vacchini@mi.infn.it}

\maketitle

\begin{history}
\received{(received date)}
\revised{(revised date)}
\end{history}

\begin{abstract}
  We consider the description of quantum noise within the framework of
  the standard Copenhagen interpretation of quantum mechanics applied
  to a composite system environment setting. Averaging over the
  environmental degrees of freedom leads to a stochastic quantum
  dynamics, described by equations complying with the constraints
  arising from the statistical structure of quantum mechanics. Simple
  examples are considered in the framework of open system dynamics
  described within a master equation approach, pointing in particular
  to the appearance of the phenomenon of decoherence and to the
  relevance of quantum correlation functions of the environment in the
  determination of the action of quantum noise.
\end{abstract}

\section{Introduction}

The corner stone for the successful description of experiments with
microscopic systems as statistical experiments was laid by Bohr through his
probabilistic reading of the square modulus of the wavefunction, finally
leading to the so called Copenhagen interpretation of quantum mechanics. This
interpretation of quantum mechanics is often also termed orthodox, to stress
the existence of alternative viewpoints, still compatible with present day
most refined experiments on the foundations of quantum mechanics (see e.g. the
special issue {\cite{NatureFoundations}}). Further developments have deepened
and strengthened the understanding of quantum mechanics as a theory describing
experiments in a statistical framework. In this spirit it has become clear
that quantum mechanics naturally leads to a new probabilistic
description with respect to the classical one, sometimes termed
quantum probability {\cite{Strocchi2005}}, so
that from now on we will use the general term quantum theory, even though it actually
started as an alternative to classical mechanics.

Quantum theory includes and extends the classical
probabilistic description, so that bringing over ideas and concepts from
classical probability theory through the quantum border is a fruitful path in
order to further understand and explore the statistical structure of quantum
theory, and \ leads to a reach variety of new phenomena (for a presentation of
quantum theory along these lines see e.g. {\cite{Holevo2001,Alicki2001}}).
Actually it is an amusing, and possibly telling, coincidence the fact that the
book in which von Neumann laid the mathematical foundations of quantum theory
{\cite{Neumann1932a}} appeared almost at the same time as the contribution in
which Kolmogorov laid the foundations of classical probability theory basing
its axiomatic presentation on measure theory {\cite{Kolmogorov1933a}}.

The Copenhagen interpretation, which tells us that the quantum description of
physical systems brings with itself an intrinsic statistical aspect, can
equally well describe composite systems, that is a situation in which one can
distinguish between different parts of the overall system. Let us call system
the subset of degrees of freedom we are interested in and can access
experimentally, as well as environment the other degrees of freedom, still to
be described with the aid of quantum theory. A relevant and interesting
question is the quantum prediction for the dynamics of the relevant degrees of
freedom we call system if one does not or can not observe the environmental
degrees of freedom. In such a situation, on top of the in principle
unavoidable statistical aspect due to the very nature of quantum theory, an
additional source of randomness appears, which can be termed quantum noise
{\cite{Gardiner2000a,Clerk2010a}}. This situation can be seen as the analogue
of what happens in a classical setting when a given system undergoes a
stochastic dynamics. However, in the classical case the dynamics of a small
isolated system is in principle deterministic, and the statistical aspect in
the description can always be seen as arising from the effect of a classical
noise, possibly effectively describing the interaction with other classical
degrees of freedom. In the quantum setting the action of quantum noise on the
contrary builds on the original statistical description. Besides this,
important constraints on the structure of the equations describing the quantum
stochastic dynamics as well as on the properties of the quantum noise itself
appear, essentially related to the non commutativity of observables, playing
the role of random variables in quantum theory, and to the tensor product
structure of the Hilbert space on which composite systems are described.

It is to be stressed that quantum noise, or actually more precisely noise in
a quantum system, can describe a phenomenon which is typical of the
quantum realm, namely decoherence. The latter can be understood as the
dynamical loss of the capability to show up quantum interference effects in a
given system basis, as a consequence of the interaction with other external
quantum degrees of freedom. We recall that the term decoherence or
dephasing is sometimes also used to describe more generically a loss
of coherence or visibility which can be obtained within a classical description.
As a result quantum noise can induce an effective classical
dynamics for certain system observables, still not selecting a definite
outcome so that it does not lead to a solution of the measurement
problem.

An interesting issue within the description of randomness in the
dynamics of a quantum system is also the distinction between noise
which can be avoided by means of a more refined control or noise which
is actually intrinsic to the quantum description
{\cite{Dhara2014a}}. Most recently an approach has also been suggested
{\cite{Arenz2015a}} in order to discriminate between decoherence arising from
an actual interaction with unobserved degrees of freedom and decoherence arising from modifications of quantum mechanics as
suggested by collapse models or other alternative theories.

In this contribution we will briefly describe the emergence of a dynamics
driven by quantum noise in the framework of open quantum system theory,
considering basic examples.

\section{Reduced system dynamics}

\begin{figure}[h]
  \centerline{\psfig{file=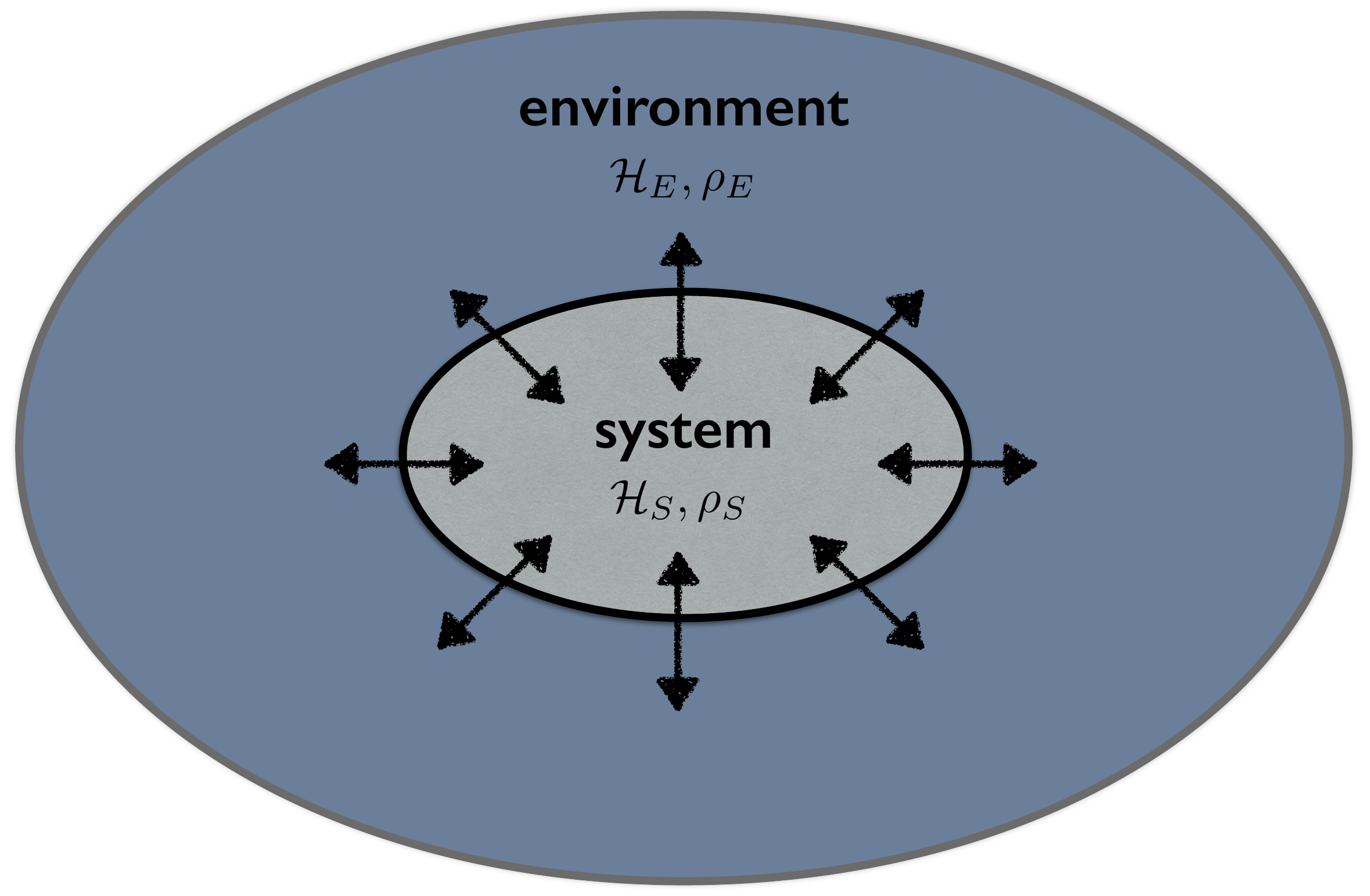,width=8cm}} {\vspace{8pt}}
  \caption{A schematic illustration of an open system with Hilbert space
  $\mathcal{H}_{S}$ and reduced state $\rho_{S}$, interacting with an
  environment described in the Hilbert space $\mathcal{H}_{E}$, with reduced
  state $\rho_{E}$.\label{fig:oqs}}
\end{figure}

Let us consider the general framework of open quantum system dynamics
{\cite{Breuer2002}}, introducing a quantum system described on the Hilbert
space $\mathcal{H}_{S}$, interacting with a quantum environment living in
$\mathcal{H_{}}_{E}$, as depicted in Fig.~\ref{fig:oqs}. If we denote with
$\rho_{SE}$ the total state and describe the interaction by means of the
unitary operators $U ( t )$ acting on $\mathcal{H}_{S} \otimes
\mathcal{H}_{E}$, further assuming that the state at the initial time is
factorized $\rho_{SE} ( 0 ) = \rho_{S} ( 0 ) \otimes \rho_{E}$, we have that
the reduced state of the system, describing the dynamics of the system's
observables only, is given by
\begin{eqnarray}
  \rho_{S} ( t ) & = & \tmop{Tr}_{E} \{ U ( t ) \rho_{S} ( 0 ) \otimes
  \rho_{E} U ( t )^{\dag} \} . \label{eq:rd} 
\end{eqnarray}
The assignment $\rho_{S} ( 0 ) \mapsto \rho_{S} ( t )$ turns out to define a
map which is in particular completely positive, that is remains positive when
extended to act on a tensor product extension of the considered Hilbert space
$\mathcal{H}_{S}$.

In many situations of interest the reduced system state dynamics is
well described by a time-local master equation of the form
\begin{eqnarray}
  \frac{\mathd}{\tmop{dt}} \rho_{S} ( t ) & = & \mathcal{L} ( t ) \rho_{S} ( t
  ) \nocomma , \label{eq:ms} 
\end{eqnarray}
where the superoperator $\mathcal{L} ( t )$ is know as generators of the
dynamics. For the case in which this superoperator is actually time
independent, according to a famous result {\cite{Lindblad1976a,Gorini1976a}}
it is known to have the so called Gorini-Kossakowski-Sudarshan-Lindblad form
\begin{eqnarray}
  \frac{\mathd}{\tmop{dt}} \rho_{S} ( t ) & = & - \frac{i}{\hbar} [ H,
  \rho_{S} ( t ) ] + \sum_{j,k} a_{jk} \left[ L_{j} \rho_{S} ( t )
  L_{k}^{\dag} - \frac{1}{2} \{ L_{k}^{\dag} L_{j} , \rho_{S} ( t ) \} \right]
  \nocomma , \label{eq:lind} 
\end{eqnarray}
where $H$ is a self-adjoint operator on the space of the system and the matrix
$a_{jk}$ has to be positive. In particular together with the system operators
$\{ L_{k} \}$ this matrix defines the details of the system-environment
interaction and depends on the quantum correlation functions of the
environment. Properties of the quantum noise affecting the open system
dynamics are therefore encoded in the operator structure of the r.h.s. of
Eq.~(\ref{eq:lind}), as well as in the features of correlation functions of
quantum operators, as we shall see in the examples. The situation described by
Eq.~(\ref{eq:lind}) corresponds to a semigroup dynamics for the open quantum
system, which can be considered Markovian, in the sense that the state of the
system at a given time is enough to determine it at later times. Actually the
proper definition of what should be considered as non-Markovian dynamics in a
quantum framework, and therefore also of non-Markovian quantum noise, has
newly become the object of an extensive research activity (see e.g.
{\cite{Rivas2014a,Breuer2016a}}). It is important to stress that the strategy
that we have here briefly outlined is certainly not the only approach to the
description of quantum noise arising within the Copenhagen interpretation by
the interaction of the system with other unobserved quantum degrees of
freedom, for the presentation of other viewpoints and techniques see
e.g.~{\cite{Alicki1987,Gardiner2000a,Weiss2008,Breuer2002}}. A crucial point
to stress is the measurement character of such time evolution, at variance
with a standard unitary dynamics. Indeed as it has been shown within the
framework of continuous measurement theory (see e.g. {\cite{Barchielli2009}}),
such a dynamics can be obtained as a result of measurements performed on the
side of the system, and can be described introducing non commuting noises. A
thorough quantum description of noise allows in particular the preservation of
basic features of quantum mechanics, such as e.g. Heisenberg's commutation
relations {\cite{Barchielli2015a}}, which are generally not accounted for in
phenomenological models which can be used to describe a stochastic dynamics.

\subsection{Decoherence models}

For the sake of example we will now briefly consider two quantum dynamics
which can be addressed within the previously introduced framework, and show
how quantum noise can induce decoherence on the system degrees of freedom,
determined by the environmental correlation functions.

Let us first consider a massive quantum particle interacting through
collisions with a background ideal quantum gas. In such a setting for a
sufficiently dilute gas memory effects can be safely neglected, so that indeed
the dynamics can be taken to be Markovian. It can therefore be assumed that
the dynamics can be described by an equation of the form Eq.~(\ref{eq:lind}),
upon suitable microscopic or phenomenological determination of the different
coefficients and operators. In this case the interaction can be naturally
taken of the form {\cite{Petruccione2005a}}
\begin{eqnarray}
  V & = & \int \mathd^{3} \tmmathbf{x} \int \mathd^{3} \tmmathbf{y}N_{S} (
  \tmmathbf{x} ) v ( \tmmathbf{x}-\tmmathbf{y} ) N_{E} ( \tmmathbf{y} ) ,
  \label{eq:v} 
\end{eqnarray}
where $N_{S} ( \tmmathbf{x} )$ and $N_{E} ( \tmmathbf{y} )$ denote the number
operator density for system and environment respectively. In this situation it
can be shown {\cite{Vacchini2009a}} that tracing over the gas degrees of
freedom the master equation takes on the form
\begin{eqnarray}
  \frac{\mathd}{\tmop{dt}} \rho_{S} ( t ) & = & - \frac{i}{\hbar} [ H_{0} ,
  \rho_{S} ( t ) ] \nonumber\\
  &  & + \int \mathd^{3} \tmmathbf{q} \mu ( \tmmathbf{q} ) \left[
  \mathe^{\frac{i}{\hbar} \tmmathbf{q} \nosymbol \cdot \hat{\tmmathbf{x}}}
  \sqrt{S ( \tmmathbf{q},E ( \tmmathbf{q}, \hat{\tmmathbf{p}} ) )} \rho_{S} (
  t ) \sqrt{S ( \tmmathbf{q},E ( \tmmathbf{q}, \hat{\tmmathbf{p}} ) )}
  \mathe^{- \frac{i}{\hbar} \tmmathbf{q} \nosymbol \cdot \hat{\tmmathbf{x}}}
  \right. \nonumber\\
  &  & \hspace{6em} \hspace{6em} \left. - \frac{1}{2} \{ S ( \tmmathbf{q},E (
  \tmmathbf{q}, \hat{\tmmathbf{p}} ) ) , \rho_{S} ( t ) \} \right] ,
  \label{eq:qbm} 
\end{eqnarray}
where $H_{0}$ is the free kinetic Hamiltonian, $\mu ( \tmmathbf{q} ) = ( 2 \pi
)^{4} \hbar^{2} n | \tilde{v} ( \tmmathbf{q} ) | ^{2}$, with $n$ gas particle
density and $\tilde{v} ( \tmmathbf{q} )$ Fourier transform of the interaction
potential. In the expression $\hat{\tmmathbf{x}}$ and $\hat{\tmmathbf{p}}$
denote position and momentum operators of the test particle, so that the
unitary operators $\mathe^{\frac{i}{\hbar} \tmmathbf{q} \nosymbol \cdot
\hat{\tmmathbf{x}}}$ describe momentum translations, while
\begin{eqnarray}
  S ( \tmmathbf{q},E ) & = & \frac{1}{2 \pi \hbar} \int \mathd t \int
  \mathd^{3} \tmmathbf{x} \mathe^{\frac{i}{\hbar} ( Et-\tmmathbf{q} \cdot
  \tmmathbf{x} )} \frac{1}{N} \int \mathd^{3} \tmmathbf{y} \langle N_{E} (
  \tmmathbf{y} ) N_{E} ( \tmmathbf{x}+\tmmathbf{y},t ) \rangle \nonumber\\
  & = & \frac{1}{2 \pi \hbar} \frac{1}{N} \int \mathd t
  \mathe^{\frac{i}{\hbar} Et} \langle \varrho^{\dag}_{\tmmathbf{q}}
  \varrho_{\tmmathbf{q}} ( t ) \rangle \label{eq:dsf} 
\end{eqnarray}
upon defining
\begin{eqnarray}
  \varrho_{\tmmathbf{q}} & = & \int \mathd^{3} \tmmathbf{x} \mathe^{-
  \frac{i}{\hbar} \tmmathbf{q} \nosymbol \cdot \tmmathbf{x}} N_{E} (
  \tmmathbf{x} ) . \label{eq:rq} 
\end{eqnarray}
The master equation is fixed by the function $S ( \tmmathbf{q},E )$ defined in
Eq.~(\ref{eq:dsf}), also known as dynamic structure factor
{\cite{Pitaevskii2003}}, which is actually the Fourier transform of the
density-density correlation function of the environment, which appears
operator-valued being evaluated in \tmtextbf{$E ( \tmmathbf{q},
\hat{\tmmathbf{p}} )$}, with $E ( \tmmathbf{q},\tmmathbf{p} ) = (
\tmmathbf{p}+\tmmathbf{q} )^{2} / ( 2M ) -\tmmathbf{p}^{2} / ( 2M )$ energy
transfer in a single collision, $M$ mass of the gas particle. Note that one
has the identity
\begin{eqnarray}
  \langle \varrho^{\dag}_{\tmmathbf{q}} \varrho_{\tmmathbf{q}} ( t ) \rangle &
  = & \frac{1}{2} \langle \{ \varrho^{\dag}_{\tmmathbf{q}} ,
  \varrho_{\tmmathbf{q}} ( t ) \} \rangle + \frac{1}{2} \langle [
  \varrho^{\dag}_{\tmmathbf{q}} , \varrho_{\tmmathbf{q}} ( t ) ] \rangle ,
  \label{eq:qcf} 
\end{eqnarray}
where the last contribution is non vanishing just due to the operator nature
of the environmental quantities in the quantum description. This quantity
depending on the density fluctuations can be directly related to the dynamic
response function of the environment $\chi^{''} ( \tmmathbf{q},E )$ according
to the fluctuation-dissipation formula
\begin{eqnarray}
  S ( \tmmathbf{q},E ) & = & \frac{1}{\pi} \frac{1}{1- \mathe^{\beta E}}
  \chi^{''} ( \tmmathbf{q},E ) . \label{eq:fdt} 
\end{eqnarray}
The considered master equation describes both dissipation and decoherence
effects in the stochastic dynamics of the particle undergoing quantum Brownian
motion. To put into evidence decoherence effects in the position
representation it is convenient to consider a simplified expression in which
we treat momentum as a classical variable, so that Eq.~(\ref{eq:qbm}) takes
the much simpler expression
\begin{eqnarray}
  \frac{\mathd}{\tmop{dt}} \rho_{S} ( t ) & = & \int \mathd^{3} \tmmathbf{q}
  \tilde{\mu} ( \tmmathbf{q} ) \left[ \mathe^{\frac{i}{\hbar} \tmmathbf{q}
  \nosymbol \cdot \hat{\tmmathbf{x}}} \rho_{S} ( t ) \mathe^{- \frac{i}{\hbar}
  \tmmathbf{q} \nosymbol \cdot \hat{\tmmathbf{x}}} - \rho_{S} ( t ) \right]
  \nocomma , \label{eq:decoh} 
\end{eqnarray}
with $\tilde{\mu} ( \tmmathbf{q} )$ a suitable positive density and its
solution in the position matrix elements can be written as
\begin{eqnarray}
  \langle \tmmathbf{x} | \rho_{S} ( t ) | \tmmathbf{y} \rangle & = & \mathe^{-
  \Lambda [ 1- \Phi ( \tmmathbf{x}-\tmmathbf{y} ) ] t} \langle \tmmathbf{x} |
  \rho_{S} ( 0 ) | \tmmathbf{y} \rangle , \label{eq:levi} 
\end{eqnarray}
with $\Phi ( \tmmathbf{x} )$ the characteristic function of the probability
distribution of momentum transfers between test and gas particles, and
$\Lambda$ a collision rate {\cite{Vacchini2005a}}. As a result off-diagonal
matrix elements in the position representation are suppressed with elapsing
time. This means in particular that if the system is initially in a coherent
superposition of spatially separated states the quantum noise can drive the
system to a classical mixture, which is a typical decoherence effect. This
kind of models can explain decoherence effects in interference experiments
with massive particles {\cite{Hornberger2003a,Hackermuller2004a}}. Note that a
similar result for the dynamics of the statistical operator $\rho_{S} ( t )$
arises in dynamical reduction models {\cite{Vacchini2007b}}, however only the
average effect can be compared, in such models one has a localization effect
acting on the single realizations, leading to a possible solution of the
measurement problem {\cite{Bassi2003a}}.

As a further example showing the relevance of quantum correlation functions in
the description of a noisy quantum dynamics we consider an exactly solvable
model of decoherence {\cite{Breuer2002}}. In this case one considers a
two-level system interacting with a bosonic reservoir according to the
coupling
\begin{eqnarray}
  V & = & \sigma_{z} \sum_{k} ( g_{k} b_{k}^{\dag} +g^{\ast}_{k} b_{k} ) ,
  \label{eq:tls} 
\end{eqnarray}
where besides the standard Pauli operator we have introduced complex coupling
coefficients $g_{k}$, as well as the creation and annihilation operators
$b_{k}$ and $b_{k}^{\dag}$ obeying the standard canonical commutation
relations. If we assume the bosonic reservoir to be in a thermal state the
reduced dynamics can be exactly worked out and leads to a master equation
which is in a form similar to Eq.~(\ref{eq:lind}), albeit with a time
dependent coefficient
\begin{eqnarray}
  \frac{\mathd}{\tmop{dt}} \rho_{S} ( t ) & = & - \frac{i}{\hbar} [ H_{0} ,
  \rho_{S} ( t ) ] + \gamma ( t ) [ \sigma_{z} \rho_{S} ( t ) \sigma_{z} -
  \rho_{S} ( t ) ] \nocomma , \label{eq:deph} 
\end{eqnarray}
where $H_{0} = \hbar \omega_{0} \sigma_{z}$ is the free system Hamiltonian.
The time dependent coefficient $\gamma ( t )$ is determined again from a
correlation function depending on the environment operators appearing in the
interaction term Eq.~(\ref{eq:tls}) and given by
\begin{eqnarray}
  \alpha ( t ) & = & \sum_{k} | g_{k} |^{2} ( \langle b_{k} ( t ) b_{k}^{\dag}
  \rangle + \langle b_{k}^{\dag} ( t ) b_{k} \rangle ) \nonumber\\
  & = & \int^{\infty}_{0} \mathd \omega J ( \omega ) \left\{ \coth \left(
  \frac{\beta}{2} \hbar \omega \right) \cos ( \omega t ) -i \sin ( \omega t )
  \right\} , \label{eq:cossin} 
\end{eqnarray}
where the two contributions at the r.h.s. come from the evaluation of the
correlation function relying on a decomposition as the one considered in
Eq.~(\ref{eq:qcf}), where now the commutator part typically related to
dissipation amounts to a $\mathbbm{C}$-number term. We have further introduced
the so called spectral density $J ( \omega ) \nocomma$, formally defined as $J
( \omega ) = \sum_{k} | g_{k} |^{2} \delta ( \omega - \omega_{k} )$, with
$\omega_{k}$ the frequency of the bosonic modes appearing in the free
Hamiltonian of the environment $\sum_{k} \hbar \omega_{k} b_{k}^{\dag} b_{k}$,
which actually allows to go over to a continuum limit embodying in itself
dependence of the coupling strength on the environment frequencies as well as
on the distribution of the environmental modes. For the decoherence dynamics
described by the present model only the anticommutator part of the correlation
function related to decoherence is relevant and one has in particular
\begin{eqnarray}
  \gamma ( t ) & = & \Re \int_{0}^{t} \mathd \tau \alpha ( t- \tau )
  \nonumber\\
  & = & \int^{\infty}_{0} \mathd \omega J ( \omega ) \coth \left(
  \frac{\beta}{2} \hbar \omega \right) \frac{\sin ( \omega t )}{\omega} .
  \label{eq:jj} 
\end{eqnarray}
In view of the interaction term Eq.~(\ref{eq:tls}) one immediately sees that
the diagonal matrix elements of the statistical operator in the basis of
eigenvectors of the system Hamiltonian are constant, while coherences are
generally suppressed according to
\begin{eqnarray}
  \langle 1 | \rho_{S} ( t ) | 0 \rangle & = & \mathe^{- \Gamma ( t )}
  \mathe^{i \omega_{0} t} \langle 1 | \rho_{S} ( 0 ) | 0 \rangle ,
  \label{eq:offd} 
\end{eqnarray}
where bra and ket denote the eigenvectors of the $\sigma_{z}$ operator, and
the decoherence function $\Gamma ( t )$ is still determined by the spectral
density and the correlation function of the environment through the expression
\begin{eqnarray}
  \Gamma ( t ) & = & \int_{0}^{t} \mathd \tau \gamma ( \tau ) \nonumber\\
  & = & \int^{\infty}_{0} \mathd \omega J ( \omega ) \coth \left(
  \frac{\beta}{2} \hbar \omega \right) \frac{1- \cos ( \omega t )}{\omega^{2}}
  . \label{eq:Gamma} 
\end{eqnarray}
As a result one has a general description of the decoherence dynamics of a
two-level system coupled to bosonic degrees of freedom, allowing for a
phenomenological modelling of the effective reservoir through the suitable
definition of a spectral density. Also in this model we have seen how the
quantum stochastic dynamics is driven by correlation functions of the
environment operators, which embody the noisy action of the quantum
environment.

\section{Conclusions and outlook}

If one considers a non isolated quantum system, its dynamics shows up an
additional layer of stochasticity, on top of the probabilistic quantum
description, which arises due to the interaction with the unobserved quantum
environmental degrees of freedom. In this perspective quantum noise can be
described applying the standard Copenhagen formulation of quantum mechanics to
the overall degrees of freedom. In this framework one is able to describe in a
consistent way both dissipative and decoherence effects. The latter lead from
a quantum probabilistic setting to a classical one, in which the interference
capability of selected quantum degrees of freedom is suppressed. As a result
one recovers a classical behaviour for certain degrees of freedom, however
still not solving the measurement problem, which has to face the fact that
macroscopic objects do appear in definite states, rather than in
superpositions or classical mixture states. Different techniques and
approaches can be devised in order to describe the quantum noisy dynamics of
such open quantum systems and in this contribution we have considered two
paradigmatic examples within the framework of a master equation approach. It
appears how the action of quantum noise in this description typically depends
on the features of two-point correlation functions of the quantum operators of
the environment involved in the interaction term.

An important open issue in this and other descriptions of quantum noise within
the standard Copenhagen interpretation is the formulation and characterization
of memory effects. Recent work on the subject {\cite{Rivas2014a,Breuer2016a}}
has put quantum non-Markovianity in connection with properties of the
statistical operator of the open system undergoing a stochastic dynamics, or
of the mapping describing the reduced dynamics. This is at variance with the
classical case, in which there is a clearcut definition of Markovian noise in
the framework of classical stochastic processes, which cannot be directly used
in the quantum framework {\cite{Vacchini2011a}}. Future characterization of
quantum noise in view of its memory properties might well be connected with
expression and features of multitime correlation functions of environmental
quantum operators. A better understanding of the description of quantum noise
will also be useful in view of a comparison between orthodox quantum mechanics
and modifications of it, such as dynamical reduction models, aiming to a
solution of the quantum measurement problem and leading to distinct
experimental predictions, which are in principle detectable. Indeed
determination of experimental bounds on the value of the parameters appearing
in such models, as well as their extension to a non-Markovian regime, are the
object of an intense research activity {\cite{Bassi2013a}}.

\section*{Acknowledgements}

The author gratefully acknowledges support from the EU Collaborative Project QuProCS
(Grant Agreement 641277) and by the Unimi TRANSITION GRANT - HORIZON 2020.

\bibliographystyle{ws-fnl}

\end{document}